\documentclass[conference]{IEEEtran}
\IEEEoverridecommandlockouts
\usepackage{graphicx}
\usepackage{cite}
\usepackage{enumerate}
\usepackage{array}
\usepackage{amsmath}
\usepackage{amssymb}
\usepackage{cases}
\usepackage{bm}
\usepackage{algorithm}
\usepackage{algorithmic}
\usepackage{multirow}

\usepackage{epsfig}

\newtheorem{theorem}{Theorem}

\newenvironment{proof}{{\bf Proof:}}{\hfill\rule{2mm}{2mm}}
\newcommand{\E}[1]{\mathbf{E}\left[#1\right]}

\def\BibTeX{{\rm B\kern-.05em{\sc i\kern-.025em b}\kern-.08em
    T\kern-.1667em\lower.7ex\hbox{E}\kern-.125emX}}

\begin{document}
\title{Context-Aware Information Lapse for Timely Status Updates in Remote Control Systems\\
\thanks{This work is sponsored in part by the Nature Science Foundation of China (No. 61871254, No. 91638204, No. 61571265, No. 61861136003, No. 61621091), National Key R\&D Program of China 2018YFB0105005, and Intel Collaborative Research Institute for Intelligent and Automated Connected Vehicles.}}
\author{\IEEEauthorblockN{Xi~Zheng, Sheng~Zhou, Zhisheng~Niu}

\IEEEauthorblockA{Beijing National Research Center for Information Science and Technology\\
Department of Electronic Engineering, Tsinghua University, Beijing 100084, P.R. China\\
zhengx14@mails.tsinghua.edu.cn,
\{sheng.zhou, niuzhs\}@tsinghua.edu.cn}}

\maketitle

\begin{abstract}
Emerging applications in Internet of Things (IoT), such as remote monitoring and control, extensively rely on timely status updates. Age of Information (AoI) has been proposed to characterize the freshness of information in status update systems. However, it only considers the time elapsed since the generation of the latest packet, and is incapable of capturing other critical information in remote control systems, such as the stochastic evolution and the importance of the source status. In order to evaluate the timeliness of status updates in remote control systems, we propose a context-aware metric, namely the context-aware information lapse. The context-aware information lapse characterizes both the stochastic evolution of the source status and the context-aware importance of the status. In this paper, the minimization of average context-aware lapse in a multi-user system is considered, and a corresponding user scheduling policy is proposed based on Lyapunov optimization. Numerical results show that compared to AoI-based policy, the context-aware-lapse-based policy can achieve a substantial improvement in terms of error-threshold violation probability and control performance.
\end{abstract}

\section{Introduction}
Applications in Internet of Things (IoT) require timely and reliable information exchange and update among various ends. For example, in vehicular networks, vehicles need to exchange position, velocity, acceleration and driving intention to enable driving assistance applications, such as collision avoidance and intersection management. 
The timeliness requirement for status updates varies according to the dynamics of the status and the context in the system. Generally speaking, when the monitored status changes rapidly (e.g., high speed or heavy brake), more frequent update of the status is required; when the monitored status is at a critical situation (e.g., collision avoidance), more information should be delivered. Otherwise, untimely status update can either be a waste of wireless resource or hinder the effectiveness of control, resulting in undesirable performance degradation. Therefore, to ensure the timeliness of information delivery and the effectiveness of control, status updates should adapt to the context of the system and the dynamics of the status. 

To cope with the changing context in the system and the stochastic evolution of the status, this paper proposes a new metric named the context-aware \emph{information lapse}. The context-aware lapse consists of two parts, namely the context-aware weight and the lapse. The lapse identifies how inaccurate the monitor's information is compared with the actual status, while the context-aware weight evaluates how crucial the status information is for decision-making in remote control. The lapse is determined by the dynamics of the status, while the context-aware weight is associated with context information in the system. 

There has been several metrics proposed to evaluate the timeliness of status updates. Age of information (AoI) \cite{aoi} is defined as the time elapsed since the generation of the most up-to-date packet received. It captures the freshness of information regardless of its context and dynamics. AoI has been studied extensively \cite{aoi}--\!\!\cite{NegExpo} in recent literatures. Especially, in \cite{8000687}--\!\!\cite{AoIEH}, the authors evaluate the penalty caused by information staleness, in order to characterize the non-linear performance of a system with respect to AoI. It is proved in many cases, such as \cite{NegExpo} and \cite{350282}, that the mean square error (MSE) of status value is a function of the AoI when the status is not observable during the scheduling of updates. However, in practical systems, the status information can be observable to the scheduler when the scheduler is close to the monitored object. In \cite{EffectiveAge}, the authors claim that minimizing AoI does not guarantee MSE minimization when the status is observable, and name two effective age metrics that are minimized when MSE is minimized. However, none of these metrics takes environmental context into consideration when designing status update schemes, which lacks insights into exploiting context information. 

In this paper, we investigate how to exploit the context information as well as the dynamics of status in status updates for remote control systems. The contributions of this paper is summarized as follows:
\begin{enumerate}
\item The context-aware timeliness requirement of status update is identified with the proposed metric, namely the  context-aware information lapse, which is the product of the context-aware weight and the lapse. Therefore, both the context information and the dynamics of the monitored object are characterized. 
\item A multi-user scheduling problem is formulated to minimize the average context-aware lapse of a status update system with multiple users, and an adaptive context-aware scheduling policy is proposed with provable performance upper bound. 
\item Simulation results show that with the proposed user scheduling policy, there is a substantial performance improvement over an AoI-based policy in the remote control problem of stabilizing CartPoles. 
\end{enumerate}

The remainder of this paper is organized as follows. Section II introduces the context-aware lapse. A multi-user scheduling problem is formulated to reduce the context-aware lapse, and a user scheduling policy is proposed in Section III. In Section IV, performance of various policies is illustrated with simulation results. Section V concludes the paper. 
                                                                                                                                                                                                                       
\section{Context-Aware Information Lapse}
The timeliness of status update in typical applications of IoT is defined by how well the status update serves the purpose of remote control. To ensure timely status updates, the information lapse, which describes the consistency of the monitor's status information with the actual status, is faced with context-dependent requirement: when the monitored object is at a critical situation, the system demands lower information lapse. Denoting the difference between the state of the monitored status and the estimated status at the monitor at time $t$ by $Q(t)$, the information lapse is denoted by $\delta(Q(t))$, where $\delta(\cdot)$ is a non-negative function (e.g., norms). The context-aware importance of status is evaluated by weight $\omega(t)$: if the context in the system presents a high requirement on information lapse, the corresponding weight $\omega(t)$ will be large. Therefore, the context-aware timeliness of status update is characterized by the lapse $\delta(Q(t))$ and the context-aware weight $\omega(t)$. 

\subsection{Definition}

To characterize the context-aware timeliness for status updates in remote control systems, we propose a new metric named the context-aware information lapse. The context-aware information lapse is defined as the product of the lapse $\delta(Q(t))$ and the context-aware weight $\omega(t)$. Mathematically, the context-aware information lapse is expressed as
\begin{equation}
\label{eqn:def}
F(t) = \omega(t)\delta(Q(t)).  
\end{equation}
Denoting the temporal derivative $\frac{\mathrm{d}}{\mathrm{d}t}Q(t)$ of error by $A(t)$, which can be \emph{negative}, the error $Q(t)$ is equivalent written as 
\begin{eqnarray}
\label{eqn:q1}
Q(t) = \int_{g(t)}^tA(t)\mathrm{d}t,
\end{eqnarray}
where $g(t)$ is the generation time of the most up-to-date packet that is received before $t$. Note that if the information lapse $\delta(Q(t))$ increases with a constant rate (i.e., $A(t) = 1$) over time and the weight $\omega(t)$ is time-invariant, the context-aware lapse is equivalent to the conventional AoI. 
Moreover, context-aware lapse can be turned into other extensions: 
\begin{enumerate}
\item If the weight $\omega(t)$ is time-invariant, mapping $\delta(\cdot)$ is linear over $\mathbb{R}_+$ and $A(t) = \frac{\mathrm{d}}{\mathrm{d}t}f\left(\Delta\left(t\right)\right)$, where function $f: \mathbb{R}_+\mapsto\mathbb{R}$ maps AoI to the system penalty, the context-aware lapse equals the non-linear AoI defined in \cite{8006543}. 
\item If the weight $\omega(t)$ is time-invariant and $\delta(x) = x^2$, the context-aware lapse tracks the squared error. 
\end{enumerate}

The discrete-time version of the context-aware lapse is
\begin{equation*}F(t) = \omega(t)\delta\left(\sum_{\tau=g(t)}^{t-1}A(\tau)\right),\end{equation*}
where $A(t)$ is the increment of error in time slot $t$. The recursive relationship is expressed as 
\begin{equation}
\label{eqn:dis}
Q(t+1) = \left(1-D(t)\right)Q(t) + A(t) + D(t)\sum_{\tau=g(t+1)}^{t}A(\tau), 
\end{equation}
where $D(t)=1$ if there is a successful status delivery in the $t$-th slot; otherwise $D(t) = 0$. 

The context-aware lapse implies how urgent the system is for a new status update, given the lapse and the context-aware weight. By reducing the context-aware lapse, the context-aware timeliness of information can be better guaranteed. 

\subsection{Analogue to Queuing}
\begin{figure}
\centering
\includegraphics[width=2in]{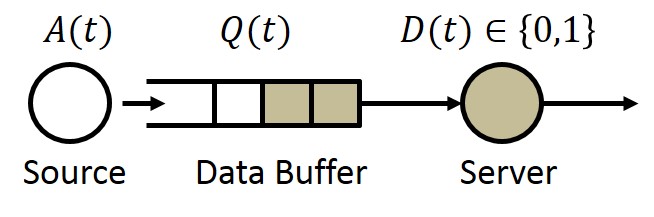}
\caption{Analogue to a queuing system in which the data buffer would be emptied if there were a delivery. }
\vspace*{-0.1in}
\label{fig:queue}
\end{figure}

Consider a special case where the delay between the generation of an update and its delivery is negligible in the system's timescale, i.e., the system state can be instantaneously obtained and transmitted by the device whenever it is scheduled. In this case, Eq. (\ref{eqn:dis}) is written as
\begin{equation}
\label{eqn:Q}
Q(t+1) = \left(1-D(t)\right)Q(t) + A(t).  
\end{equation}

The dynamic function of $Q(t)$ in (\ref{eqn:Q}) is equivalent to the queuing system in Fig. \ref{fig:queue}, where there is a source generating $A(t)$ packets at time $t$, a data buffer to store the generated packets before they are served, and a server. A major difference between Fig. \ref{fig:queue} and conventional queuing is that once the server completes a service, i.e., $D(t)=1$, the data buffer will be emptied. Another difference is that both $A(t)$ and $Q(t)$ can be negative. Nonetheless, the methods in queuing problem can be applied to the analysis of context-aware lapse. 

\section{Multi-User Scheduling}

\begin{figure}
\centering
\includegraphics[width=2in]{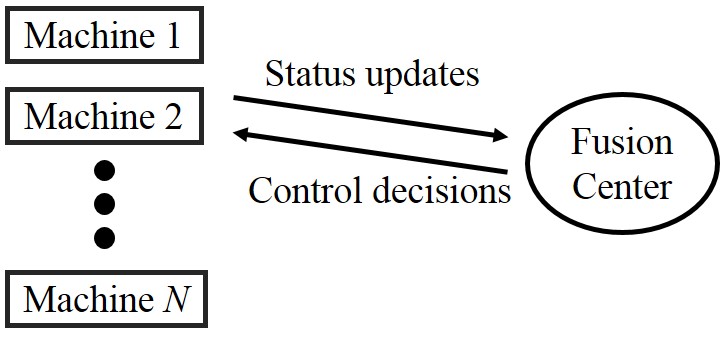}
\caption{A remote control system where a fusion center controls several machines based on their status information. }
\vspace*{-0.08in}
\label{fig:control}
\end{figure}

Consider a system with $N$ users and one fusion center, as illustrated in Fig. \ref{fig:control}. The users need to deliver their status information to the fusion center to update their status. However, due to limited channel resource, at each time slot, at most $K<N$ users can transmit simultaneously. The decision of user scheduling at the $t$-th time slot is denoted by a vector $\bm{U}(t) = \left(U_1(t),U_2(t),\cdot,U_N(t)\right)$, where $U_i(t)=1$ represents that the $i$-th user is scheduled at the $t$-th slot; otherwise $U_i(t)=0$. The updates of each user are transmitted through a block fading channel. It is assumed that the probability of successfully delivering the $i$-th user's information to the fusion center (i.e., the channel is good) is $p_i$. The state of the $i$-th user's channel being good at the $t$-th slot is represented by $S_i(t)=1$; otherwise $S_i(t)=0$. Therefore, there is a successful delivery for the $i$-th user at the $t$-th time slot \emph{if and only if} the update indicator $D_i(t) = U_i(t)S_i(t)=1$. 

Due to the randomness in status evolution, the error at the monitor might decrease even if there is no successful delivery. Therefore, it is assumed that the increment $A_i(t)$ of error can be negative, and has zero mean and variance $\E{A_{i}^2}$. It is further assumed that $A_i(t)$ is independent of the current error $Q_i(t)$. An example that satisfies these assumptions is when $A_i(t)$ are independent and identically distributed (i.i.d.) Gaussian random variables. The time-variant context-aware weight $\omega_i(t)$, which is associated with stochastic context information, is assumed to be a random variable that has mean $\E{\omega_{i}}$, and is independent of error $Q_i(t)$. 

To improve the timeliness of status updates, the average context-aware lapse minimization problem is formulated as
\begin{align}
\label{program:main}
&\min	~&& \lim_{T\to\infty}\frac{1}{T}\E{\sum_{t=0}^{T-1}\sum_{i=1}^{N} \omega_i(t)Q_i(t)^2}\\
&\mathrm{s.t.}	~&& \sum_{i=1}^N U_i(t) \leq K, \forall t \in \mathbb{N},\notag
\end{align}
where the lapse we considered here is the squared error, and the constraint corresponds to the limitation of channel resource. 
According to Eq. (\ref{eqn:Q}), the transmission at the $t$-th time slot starts to take effect at the next slot. Therefore, to reduce the context-aware lapse according to context information, we need to foresee future weight $\omega(t+1)$; otherwise problem (\ref{program:main}) is equivalent to a context-unaware problem. The prediction of future weight is quite reasonable in practical applications, since the length of a time slot in wireless communications is relatively small compared to the timescale of the context.  

By carefully designing scheduling decision $\bm{U}(t)$ at each time slot, we try to reduce the average context-aware lapse to improve the performance of the status update system. We obtain a scheduling policy by Lyapunov optimization: 

\begin{align}
\label{p:main}
&\max~ && \sum_{i=1}^N\left(\E{\omega_i}\left(\frac{1}{p_i\pi_i}-1\right)+\omega_i(t+1)\right)\times\notag\\
&~&&~~~~~~p_iQ_i(t)^2U_i(t)\\
&\mathrm{s.t.}&&\sum_{i=1}^N U_i(t) \leq K, \notag
\end{align}
where $\pi_i$ is the probability of scheduling the $i$-th user with a randomized stationary policy $\bm{\pi} = \left(\pi_1,\pi_2,\cdots,\pi_N\right)$ that satisfies $\pi_i\leq1$ and $\sum_{i=1}^N\pi_i\leq K$. An equivalent description of policy (\ref{p:main}) is to schedule $K$ users with the largest value of
$$\left(\E{\omega_i}\left(\frac{1}{p_i\pi_i}-1\right)+\omega_i(t+1)\right)p_iQ_i(t)^2.$$

\begin{theorem}
In a status update system with $N$ users and $K$ orthogonal wireless channels, the average context-aware lapse under policy (\ref{p:main}) is upper bounded by
\begin{eqnarray}
\lim_{T\to\infty}\frac{1}{T}\sum_{t=0}^{T-1}\sum_{i=1}^N\omega_{i}(t)Q_{i}(t)^2\leq\sum_{i=1}^N\frac{\E{\omega_i}}{p_i\pi_i}\E{A_i^2}. 
\end{eqnarray}
\begin{proof}
See Appendix A. 
\end{proof}
\end{theorem}

\begin{figure}
\centering
\includegraphics[width=2.3in]{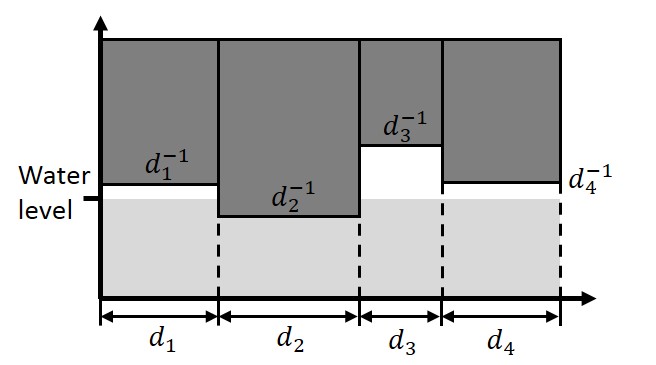}
\caption{Water-filling algorithm to obtain the parameters $\pi_i$ when $K>1$. }
\label{fig:wf}
\end{figure}

Next, we try to minimize the upper bound by selecting an appropriate randomized stationary policy $\bm{\pi}$. The problem becomes
\begin{eqnarray}
\label{pro:pi}
\begin{aligned}
&\min_{\bm{\pi}} ~~&& \sum_{i=1}^N\frac{\E{\omega_i}\E{A_i^2}}{p_i\pi_i}\\
&\mathrm{s.t.}~~&&\sum_{i=1}^N \pi_i \leq K, \pi_i \in [0,1].
\end{aligned}
\end{eqnarray}
If only one user can be scheduled at each time slot, i.e., $K = 1$, the solution to problem (\ref{pro:pi}) is 
\begin{eqnarray*}
\pi_i = \frac{\sqrt{\frac{\E{\omega_i}\E{A_i^2}}{p_i}}}{\sum_{i=1}^N\sqrt{\frac{\E{\omega_i}\E{A_i^2}}{p_i}}}. 
\end{eqnarray*}
If $K > 1$, the Karush-Kuhn-Tucher (KKT) condition is 
\begin{eqnarray*}
\left\{
\begin{aligned}
&\lambda_0\left(\sum_{i=1}^N \pi_i - K\right) = 0;\\
&\lambda_i\left(\pi_i-1\right) = 0;\\
&-\frac{\E{\omega_i}\E{A_i^2}}{p_i\pi_i^2} + \lambda_0 + \lambda_i = 0. 
\end{aligned}
\right.
\end{eqnarray*}
The solution to problem (\ref{pro:pi}) is equivalent to the water-filling problem in Fig. \ref{fig:wf}, where the total volume of the water is $K$, the width of each bar is $d_i=\sqrt{\frac{\E{\omega_i}\E{A_i(t)^2}}{p_i}}$, the height of each bar is $d_i^{-1}$, such that the maximum volume of water in each bar is $1$. The volume of water in each bar gives the probability $\pi_i$ in the randomized stationary policy $\bm{\pi}$. 

With policy (\ref{p:main}), the user with higher transmission success probability and higher context-aware weight will be granted higher priority. A larger lapse also leads to a higher priority. 

\section{Numerical Results}

\begin{figure}[t]
\centering
\includegraphics[width=2in]{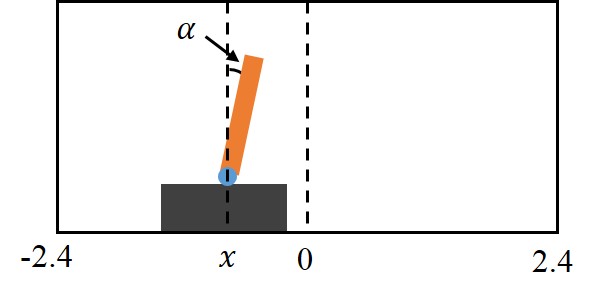}
\caption{An illustration of Cartpole. The controller pushes the cart to the left or right with 10N of force to keep the pole stand and the cart in the screen. }
\label{fig:cartpole}
\end{figure}

To evaluate the effectiveness of the proposed metric and the corresponding scheduling policy, we compare the performance of several scheduling policies in terms of average context-aware lapse, threshold violation probability, and their performance in \emph{CartPole}. In CartPole, a pole is attached to a cart (as shown in Fig. \ref{fig:cartpole}), and a controller forces the cart to its left or right to prevent the pole from falling and the cart from moving out of the screen. The game ends when the angle of the pole is larger than $12^\circ$ or the position of the cart is 2.4 units away from the center, and is played for at most 200 steps in each episode. The controller tries to play as many steps as possible in each episode. Detailed description about the game can be found in \cite{CartPole}. In the original CartPole, the 4-dimensional system status (i.e., the position $x$ and velocity $\dot{x}$ of the cart, and the angle $\alpha$ and the angular velocity $\dot{\alpha}$ of the pole) is fully observable at each step, and the controller decides whether to force the cart to its left or its right based on the status. To further complicate the CartPole game in the simulation, an additional random force is applied to the cart at each step, so that the status of the CartPole is not predictable to the controller even if the kinetics equation is known. The simulation process is described as follows:
\begin{enumerate}
\item Train a multi-layer perceptron (MLP) with a 100-node hidden layer for the control problem of a single CartPole whose the status is observable. 
\item At each step, schedule the CartPoles to deliver their status to the controller according to a scheduling policy. For the CartPoles whose status is not delivered, the controller updates their status with previous information and the kinetics equation assuming zero random force. 
\item The controller remotely controls the CartPoles with the control algorithm based on the latest available status. 
\end{enumerate}

In the simulation, the status update system has $N=10$ users (CartPoles) who share $K=2$ orthogonal wireless channels. The transmission success probability of each user is an arithmetic sequence from $0.9$ to $1$. Five polices are compared in the simulation. They are:
\begin{enumerate}
\item Round robin: Users are scheduled one by one. 
\item AoI based: Schedule $K$ users with the largest $$p_i\Delta_i(t)(\Delta_i(t)+1),$$where $\Delta_i(t)$ denotes the AoI of the $i$-th user at time $t$. Ref. \cite{Jiang} shows that the policy is asymptotically AoI-optimal when channel failure probability goes to zero and the number of users is large. 
\item AoI and context-aware weight: In many cases, the scheduler does not have access to the real-time lapse in each user node and schedule the users based on only the AoI $\Delta_i(t)$ and the weight $\omega_i(t)$. To obtain a scheduling policy without lapse, we replace $Q_i(t)^2$ with its expected value $\Delta_i(t)\E{A_i^2}$ in policy (\ref{p:main}), and the policy becomes scheduling $K$ users with the largest value of
\begin{eqnarray*}
\left(\frac{1}{p_i\pi_i}-1+\frac{\omega_i(t+1)}{\E{\omega_i}}\right)~p_i\E{\omega_i}\E{A_i^2}\Delta_i(t).
\end{eqnarray*}
\item Information lapse: If the context-aware weight at the next time slot is not observable, we take expectation over the weight in (\ref{p:main}), and obtain policy that does not require the knowledge of real-time context-aware weight, which is scheduling $K$ users with the largest $\frac{\E{\omega_i}}{\pi_i}Q_i(t)^2$. 
\item Context-aware information lapse: The context-aware scheduling policy with the knowledge of the context information at the next time slot, as is described in (\ref{p:main}). 
\end{enumerate}

\begin{figure}[t]
\centering
\includegraphics[width=2.9in]{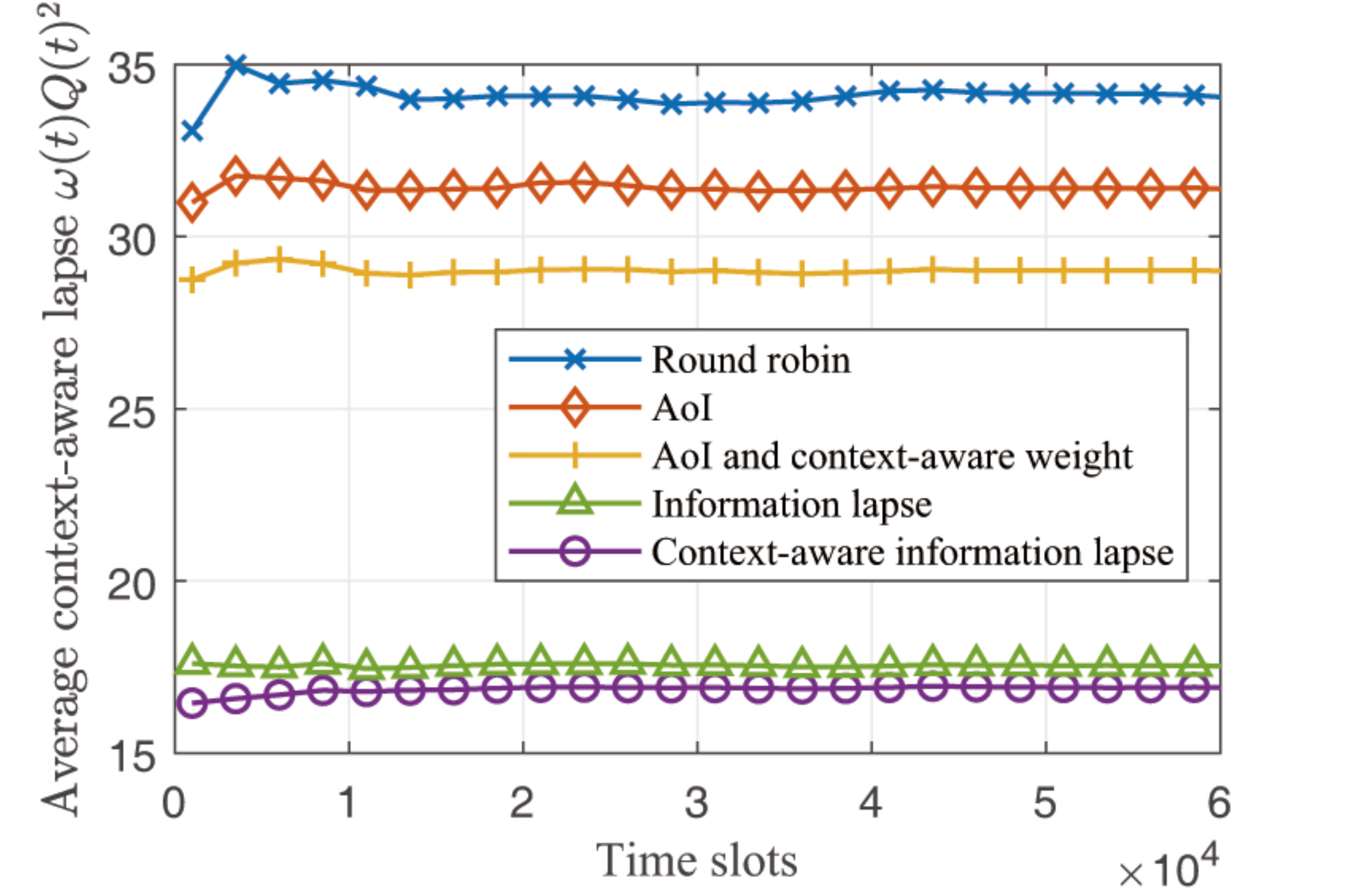}
\caption{Average context-aware lapse under several scheduling schemes. }
\label{fig:SAoI}
\end{figure}

In the first simulation, the increment of error $A_i(t)$ are i.i.d. random variables following the standard Gaussian distribution, while the context-aware weight is i.i.d. random variable with 5\% being 9 and 95\% being 1. The average context-aware lapse under the five policies is illustrated in Fig. \ref{fig:SAoI}. With both context information and information lapse, policy (\ref{p:main}) produces the lowest average context-aware information lapse. The round robin policy, which exploits nearly zero information in the system, has the highest context-aware information lapse. The performance of the AoI-based policy and the context-aware AoI policy has a significant gap, which implies that knowledge of the context-aware weight is beneficial to the scheduling. However, the difference between the performance of information lapse policy and context-aware information lapse policy is relatively smaller. The reason can be that the knowledge of one-step-ahead context-aware weight cannot greatly reduce the expected context-aware information lapse. To further improve the performance, the prediction of more further context is necessary. 

\begin{figure}[t]
\centering
\includegraphics[width=3in]{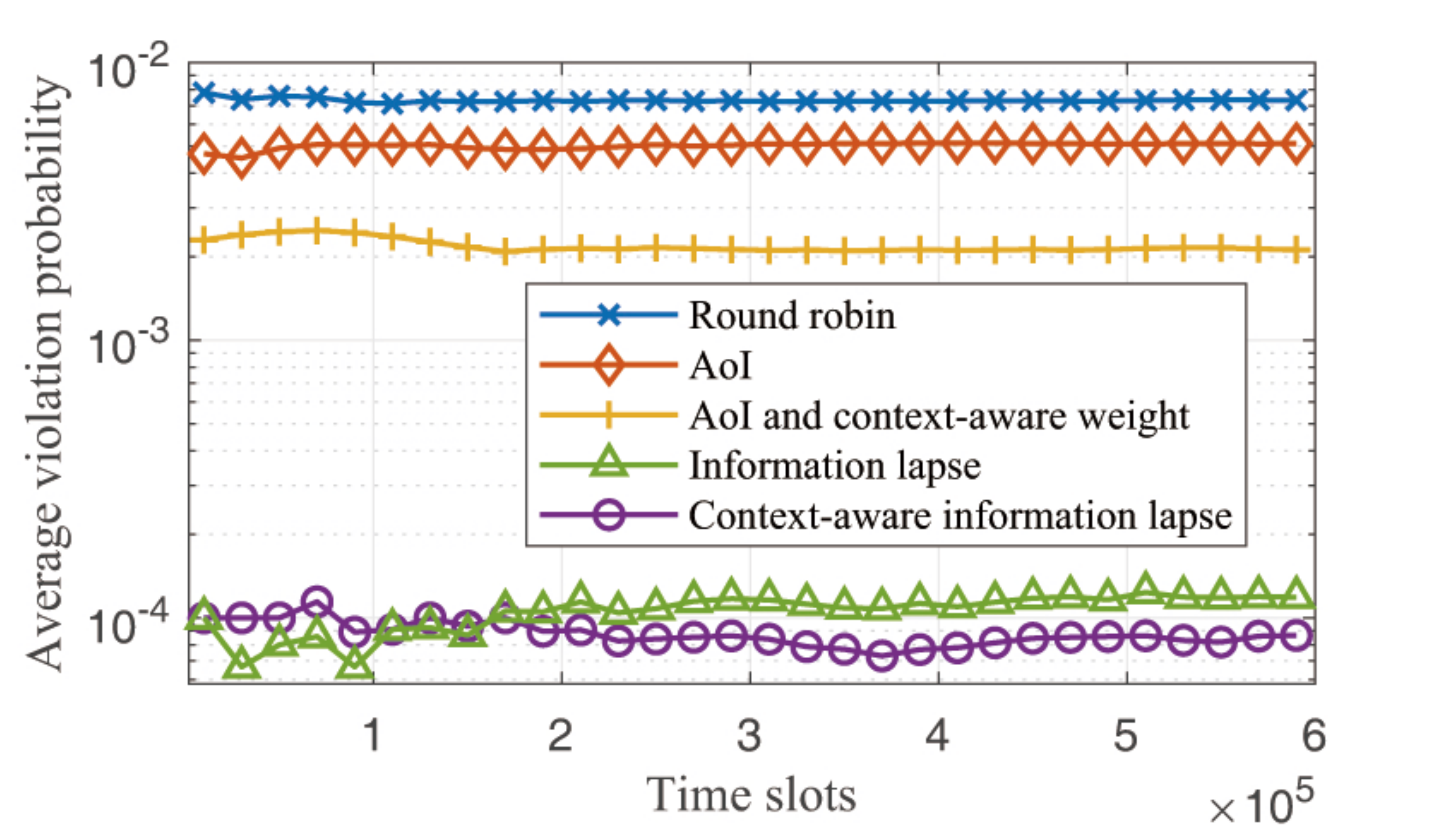}
\caption{Average threshold violation probability. The threshold for error $Q_i(t)$ is $15$ when the weight is 1, and 5 when the weight is 9. }
\label{fig:vio}
\end{figure}

Fig. \ref{fig:vio} plots the average probability of the error $|Q_i(t)|$ exceeding threshold 5 when the context-aware weight is 9 and threshold 15 when the context-aware weight is 1. It is shown that when context-aware information lapse is known to the scheduler, policy (\ref{p:main}) has a violation probability of approximately $10^{-4}$, which is the lowest of all and is approximately $\frac{1}{50}$ of the violation probability under the AoI based policy. Although the context-aware lapse minimization problem is not specifically designed for avoiding threshold violation, simulation results show that policy (\ref{p:main}) brings a significant improvement over other user scheduling policies. 

\begin{figure}[t]
\centering
\includegraphics[width=3in]{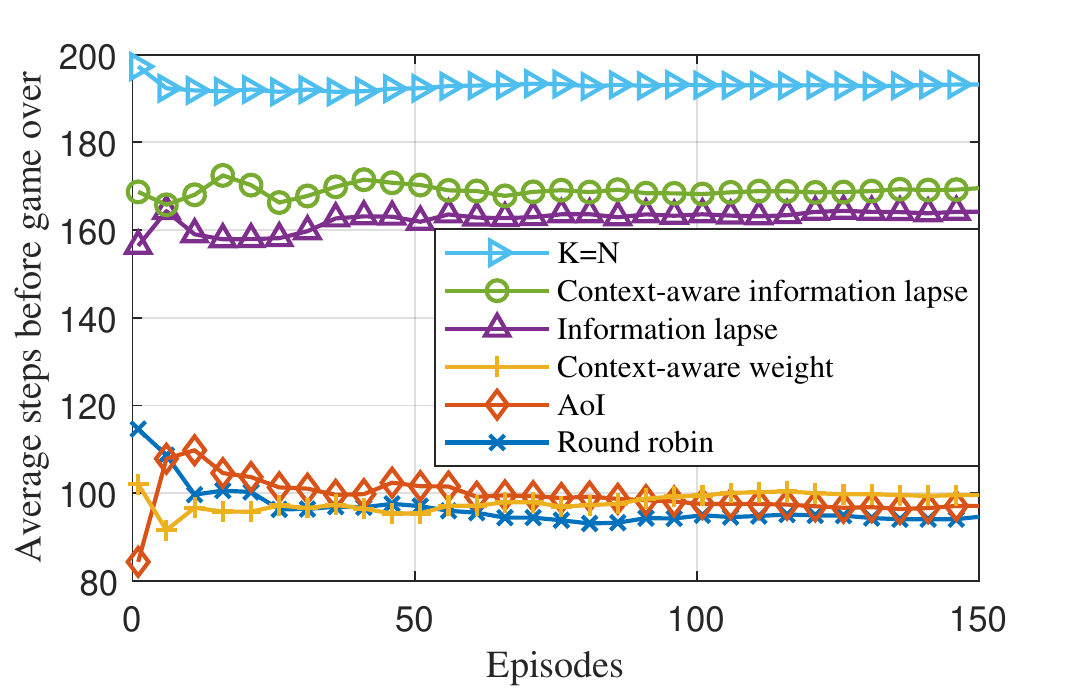}
\vspace*{-0.1in}
\caption{Average steps before game over in CartPole. }
\label{fig:ctrl}
\vspace*{-0.1in}
\end{figure}

Fig. \ref{fig:ctrl} illustrates the average number of steps being played in each episode of CartPole. The game runs on OpenAI gym\footnote{https://github.com/openai/gym/blob/master/gym/envs/classic\_control\\/cartpole.py}. The random force follows Gaussian distribution with zero mean and standard deviation being $10$ Newtons, which is the magnitude of the force applied by the controller. The context information is exploited under a straightforward intuition: the status information is more urgent if the situation worsens. Therefore, if the distance is going to increase at the next step (i.e., $x\dot{x} > 0$), the context-aware weight for position $x$ and velocity $\dot{x}$ is set to 9; otherwise the context-aware weight is 1. Similarly, the context-aware weight for angle $\alpha$ and angular velocity $\dot{\alpha}$ is 9 if $\alpha\dot{\alpha} > 0$. For each dimension of the status $(x,\dot{x},\alpha,\dot{\alpha})$, the context-aware lapse is computed. After linearly rescaling the status such that both the value of position and the value of angle are within $[-1,1]$, the context-aware lapse of each dimension is summed up as the overall context-aware lapse. Fig. \ref{fig:ctrl} shows that with the information lapse known to the scheduler, the performance of the CartPole game can be substantially improved. Additional context information, even with such simple setting of the weight, can further increase the number of steps played in each episode. 

\section{Conclusions}
This paper introduces the concept of context-aware information lapse to characterize the context-aware timeliness of a status update system. The context-aware information lapse embeds the context information in status update as well as the change of the status itself. A scheduling policy is proposed to allocate channel resources to multiple users in a multi-user status update system by reducing the context-aware lapse. Simulation results show that the proposed metric and policy achieve a significant improvement over existing AoI-based schemes on the timeliness of status updates in remote control. 

\begin{appendices}
\section{Proof for Theorem 1}
The Lyapunov function is defined as $$L(t) =  \sum_{i=1}^N\theta_iQ_i(t)^2, $$
where $\theta_i$ is a constant weight associated with the $i$-th queue. The conditional Lyapunov drift is expressed as \begin{equation}\label{eqn:drift}
\Delta(t) = \E{L(t+1) - L(t)|\bm{Q}(t), \bm{\omega}(t+1)}. 
\end{equation}
Let $f(t)$ denote the additive penalty that occurs at the $t$-th time slot. Substituting Eq. (\ref{eqn:Q}) into Eq. (\ref{eqn:drift}), we have the drift plus penalty
\begin{eqnarray*}
&~&\Delta(t) + \E{f(t)|\bm{Q}(t), \bm{\omega}(t+1)}\\
&=& \E{f(t)|\bm{Q}(t), \bm{\omega}(t+1)} + \sum_{i=1}^N\theta_i\E{A_i^2} \\
&~& - \sum_{i=1}^N\theta_ip_iQ_i(t)^2\E{U_i(t)|\bm{Q}(t), \bm{\omega}(t+1)}.
\end{eqnarray*}
Letting 
\begin{equation}
\label{eqn:penalty}
f(t) = \sum_{i=1}^N\omega_{i}(t+1)Q_{i}(t+1)^2 - \sum_{i=1}^N\omega_{i}(t+1)A_i^2(t), 
\end{equation}
the drift plus penalty becomes
\begin{align*}
&~&&\Delta(t) + \E{f(t)|\bm{Q}(t), \bm{\omega}(t+1)}\\
&=&&  - \sum_{i=1}^N\left(\theta_i+\omega_i(t+1)\right)p_iQ_i(t)^2\E{U_i(t)|\bm{Q}(t), \bm{\omega}(t+1)}\\
&~&& + \sum_{i=1}^N\theta_i\E{A_i^2} + \sum_{i=1}^N\omega_{i}(t+1)Q_i(t)^2.
\end{align*}
At each step, we try to minimize the drift plus penalty by scheduling, i.e., by choosing the appropriate $\bm{U}(t)$ to minimize the last two terms at the right-hand side of the above equality. The policy is written as:  
\begin{eqnarray}
\label{policy:0}
\begin{aligned}
&\max && \sum_{i=1}^N\left(\theta_i+\omega_i(t+1)\right)p_iQ_i(t)^2U_i(t)\\
&\mathrm{s.t.}&&\sum_{i=1}^N U_i(t) \leq K. 
\end{aligned}
\end{eqnarray}

Let $\bm{\pi} = \left(\pi_1,\pi_2,\cdots,\pi_N\right)$ be a randomized stationary policy with which the $i$-th user is scheduled with probability $\pi_i$ at each time slot. Since the drift plus penalty at each time slot is minimized by the policy (\ref{p:main}), we have that 
\begin{eqnarray*}
&~&\E{L(t+1) - L(t) + f(t)|\bm{Q}(t), \bm{\omega}(t+1)}\\
&\leq& - \sum_{i=1}^N\left(\theta_ip_i\pi_i-\omega_i(t+1)(1-p_i\pi_i)\right)Q_i(t)^2 \\
&~& + \sum_{i=1}^N\theta_i\E{A_i^2}.
\end{eqnarray*}
Since $\bm{\omega}(t+1)$ is independent to $\bm{Q}(t)$, by taking expectation and letting $\theta_i = \frac{\E{\omega_i}(1-p_i\pi_i)}{p_i\pi_i},$ we have
\begin{eqnarray}
\label{eqn:lya}
&~&\E{L(t+1) - L(t) + f(t)|\bm{Q}(t)}\notag\\
&\leq& \sum_{i=1}^N\frac{\E{\omega_i}(1-p_i\pi_i)}{p_i\pi_i}\E{A_i^2}, 
\end{eqnarray}
where the right-hand side is constant. Taking expectation at the both sides of Eq. (\ref{eqn:lya}), summing over $t\in\{0,1,\cdots,T-1\}$, dividing both sides by $T$, and taking limit $T\to\infty$, we have 
\begin{eqnarray*}
\lim_{T\to\infty}\frac{1}{T}\sum_{t=0}^{T-1}\E{f(t)}\leq\sum_{i=1}^N\frac{\E{\omega_i}(1-p_i\pi_i)}{p_i\pi_i}\E{A_i^2}. 
\end{eqnarray*}
By Eq. (\ref{eqn:penalty}), we obtain the upper bound of average context-aware lapse under policy (\ref{policy:0}):
\begin{eqnarray*}
\lim_{T\to\infty}\frac{1}{T}\sum_{t=0}^{T-1}\sum_{i=1}^N\omega_{i}(t)Q_{i}(t)^2\leq \sum_{i=1}^N\frac{\E{\omega_i}}{p_i\pi_i}\E{A_i^2}.
\end{eqnarray*}
With $\theta_i = \frac{\E{\omega_i}(1-p_i\pi_i)}{p_i\pi_i}$, policy (\ref{policy:0}) is equivalent to policy (\ref{p:main}). Therefore, the proof is completed. 
\end{appendices}

\bibliographystyle{ieeetr}

\end{document}